\documentclass[prl,aps,epsfig,twocolumn,showpacs]{revtex4}
\usepackage{epsfig,amssymb}

\newcommand\beq{\begin{equation}}
\newcommand\eeq{\end{equation}}
\newcommand\bea{\begin{eqnarray}}
\newcommand\eea{\end{eqnarray}}

\newcommand\bib{\bibitem}

\newcommand\bi{\begin{itemize}}
\newcommand\ei{\end{itemize}}

\begin{document}

\textheight=23.8cm

\title{\Large Non-linear sigma model approach to quantum 
spin chains }
\author{\bf Sumathi Rao} 
\affiliation{\it 
Harish-Chandra Research Institute, Chhatnag Road, Jhusi, Allahabad 
211019, India\\
email address : sumathi@mri.ernet.in}

\date{\today}

\begin{abstract}
We introduce and motivate the study of quantum spin chains on a
one-dimensional lattice. We classify the varieties of methods that have
been used to study these models into three categories, - \\a) exact
methods to study specific models b) field theories to describe
fluctuations about the classical ordered phases and c) numerical methods.
We then discuss the $J_1$-$J_2$-$\delta$ model in some detail and end with a
few comments on open problems.  
\end{abstract}
\maketitle
\vskip .6 true cm

\section{\bf I. Introduction}

We start with the definition of a spin chain\cite{haldane,affleck} 
as a spin model on a
one-dimensional lattice - $e.g.$, 
\beq
H = J_1 \sum_{nn} {\bf S}_i \cdot {\bf S}_j +
J_2 \sum_{nn} ({\bf S}_i \cdot {\bf S}_j)^2  + 
J_3 \sum_{nnn} {\bf S}_i \cdot {\bf S}_j + \dots
\eeq
Here, $i,j$ represent the sites on a lattice and and the notation $nn
~(=<i,j>)$ stands
for nearest neighbour, $nnn$ stands for next nearest neighbour and so on.
The spins are Heisenberg spins satsifying
$
[{\bf S}_i^a,{\bf S}_i^b] 
= i \epsilon^{abc} {\bf S}_i^c
$
and not classically commuting variables, and hence it is a quantum spin chain.
We would like to find the ground state and excitation spectrum of these
models.

But why are we interested in these models? Spin systems as models of
magnetic materials have been used for many years\cite{textbook}
because there exist
large classes of materials where the electron stays localised and magnetic
properties reside in the individual atoms - $i.e.$, one has localised moments
which can be modelled by the spins. 

But more specifically, there are several
reasons for studying one-dimensional quantum spin chain. The first is simply
that there really exist materials that behave like one-dimensional
antiferromagnets\cite{expts,moreexpts}. 
$CsNiCl_3$ is one of them, because the ratio between the
intra-chain coupling and inter-chain coupling in this material is 0.018.
Another compound which is even more markedly one-dimensional is NENP
($Ni(C_2H_8N_2)_2(N0_2)Cl O_4$) where the ratio is of the order $10^{-4}$.
In both these materials, a gap in the excitation spectrum was found although
translational symmetry remained unbroken. This was an experimental
verification of a conjecture by Haldane\cite{haldane,affleck, revs} 
that $S=1$ Heisenberg antiferromagnets
should have a gap in the spectrum (unlike $S=1/2$) and would not break
translational symmetry (unlike dimers). 
More recently, even more 
exotic compounds which are quasi-one-dimensional and can be modelled
by  unusual spin chains (sawtooth spin chains) with missing bonds $viz$, $
H = J\sum_{i}
{\bf S}_i \cdot {\bf S}_{i+1} + J/2\sum_i(1+(-1)^i)
{\bf S}_i \cdot {\bf S}_{i+2}
$
have been found\cite{sawtooth}.

The second reason is that there exist exact solutions of some toy models,
which can then be used as a check or testing ground for new anlytical or
numerical methods.  Finally, quantum anti-ferromagnets in higher dimensions
have become particularly prominent in the last few years in the context of
high $T_c$ superconductors. It is hoped that some of the methods to solve
quantum spin chains  may have generalisation to higher dimensions.

\section{\bf II. Varieties of approaches to solve quantum spin chains}

In this section, we will discuss the various methods that have been
used to `solve' models of quantum spin chains. 

\subsection{1. Spin-wave theory} 

In higher dimensions, the
standard way to proceed is to start with the classical ground state and
then use spin-wave theory. We first  try to apply that method to the
one-dimensional spin models here. Let us start with the simplest spin-chain,
the Heisenberg antiferromagnet ($HAFM$),
\beq
H = J\sum_i {\bf S}_i {\bf S}_{i+1}~.
\eeq  
Here, $i$ runs over the sites on the one-dimensional lattice.
If the spins were classical vectors, then 
\beq
H = JS^2 \sum_i \cos~(\theta_i - \theta_{i+1})
\eeq
which is obviously minimum when $\cos (\theta_i - \theta_{i+1})= -1
\Longrightarrow (\theta_i - \theta_{i+1}) =\pi$.

Hence, the classical ground state ($Neel$ state) is given by 
\beq
|s,-s,s,-s,\dots > = | \uparrow,\downarrow,\uparrow,\downarrow,\dots>.
\eeq
Note that this is not an eigenstate of the Hamiltonian, because terms
in the Hamiltonian flips nearest neighbour spins. However, for very
large spins
\beq
[{\bf S}_i^a,{\bf S}_i^b] =\epsilon_{abc} {\bf S}_i^c =O(S) \ll O(S^2).
\eeq
Hence, in the limit ${ S} \rightarrow \infty$, the $Neel$ state must be the
ground
state. By perturbing about the $Neel$ state, we can get the results for large
but finite spin. This perturbation theory  is called the spin-wave theory and
is done using the Holstein-Primakoff transformation\cite{textbook}, 
which is given by
\bea
S_i^z = S-a_i^\dagger a_i &,& S_i^z = -S+b_i^\dagger b_i \nonumber \\
S_i^+ = \sqrt{2} S (1-{a_i^\dagger a_i \over 2S})^{1/2} a_i &,&
S_i^+ = \sqrt{2} S b_i^\dagger (1-{b_i^\dagger b_i \over 2S})^{1/2} \nonumber
\\ 
S_i^- = \sqrt{2} Sa_i^\dagger (1-{a_i^\dagger a_i \over 2S})^{1/2}  &,&
 S_i^- = \sqrt{2} S (1-{b_i^\dagger b_i \over 2S})^{1/2} b_i \nonumber \\
\eea
for the $A$ and $B$ sub-lattices, which are denoted as $i \in 
A$ when $i$ is even and  $i \in
B$ when $i$ is odd or vice-versa. We can easily check that the spins
satisfy the spin algebra when the $a_i,b_i$ and their conjugates
satisfy bosonic commutation relations. Note that in the $A$ sublattice, the
absence of any bosons in a state implies that it has the maximum spin and 
for the $B$ sub-lattice, the absence of any bosonic excitation 
 implies minimum spin. 
In the large $S$ limit, the awkward
square-root term can be dropped and the spin raising and lowering operators 
can be approximated merely as 
\bea
S_i^+ \rightarrow \sqrt{2} S a_i^\dagger&,& S_i^- \rightarrow \sqrt{2}  a_i
\nonumber  \\
S_i^+ \rightarrow \sqrt{2} S b_i^\dagger&,& S_i^- \rightarrow \sqrt{2}  b_i
\eea
on the $A$ and $B$ sub-lattices. In fact, we can develop a systematic
$1/S$ expansion by expanding the square-root term, with the above terms being
the first in the expansion. But in this review, we will stop with 
the first term.
Next, we write the Hamiltonian in terms of these bosons (using the above
approximation) as
\beq
H = J\sum_{<i,j>} [ -S^2 + S(a_i^\dagger a_i +b_j^\dagger b_j + a_ib_j
+a_i^\dagger b_j^\dagger)] 
\eeq
After going to momentum space and performing a Boguliobov  transformation, we
get 
\beq
H = \sum_{{k} \in RBZ}E_{k} (c_{ k}^\dagger c_{k} +
d_{ k}^\dagger d_{k})
\eeq
with  $E_{k} = 2JS \sin |{k}|$. As ${k}\longrightarrow 0$,
$E_{k} \longrightarrow 2JS|{k}|$, which implies that the 
$c$ and $d$ bosons, which are the spin-wave modes, are massless and
relativistic modes with spin-wave velocity given by $v_s=2JS$. This,
in fact, gives us a clue that a relativistic field theory description of the
spin-wave modes might be possible.

We can also understand more physically why there are two massless spin-wave
modes. The $Neel$ state breaks the $SO(3)$ symmetry of the spin variables down
to $SO(2)$ (rotations about the $S^z$ axis). The spin-waves are the Goldstone
modes of this spontaneous symmetry break down. ( Choosing a direction for the
$Neel$ state (ground state) spontaneously breaks the $SO(3)$ spin symmetry ot
the Hamiltonian down to $SO(2)$).

Spin-wave theory works quite well for three dimensional magnets, 
but in low dimensions, spin-wave theory has problems due to quantum
fluctuations. Let us calculate the reduction in the sub-lattice magnetisation
due to quantum fluctuations ( in arbitrary dimensions).
This can be done by computing the expectation
value of $<S_{\bf i}^z>$.    
\beq
<S_i^z> = <S-a_{\bf i}^\dagger a_{\bf i}> = S -<\sum_{\bf k} 
a_{\bf k}^\dagger  a_{\bf k}>
\eeq
In terms of the spin-wave modes, this can be rewritten as 
\bea
<S_i^z> = S &-& \sum_{\bf k}[ |u_{\bf k}|^2 <c_{\bf k}^\dagger c_{\bf k}>
+ u_{\bf k}^* |v_{\bf k} <c_{\bf k}^\dagger d_{\bf k}^\dagger> \nonumber\\
&+&|v_{\bf k}|^2 <d_{\bf k}^\dagger d_{\bf k}> 
+u_{\bf k} |v_{\bf k}^* <c_{\bf k} d_{\bf k}> \nonumber \\
&+&|v_{\bf k}|^2 ]~.
\eea
All the expectation values are zero in the ground state and we are left
with
\beq
S_{\bf i}^z = S- \sum_{\bf k}|v_{\bf k}|^2 \sim S - \int {d^d k\over (2\pi)^d}
{1\over k},
\eeq
which is linearly divergent in one dimension and logarithmically divergent in
two dimensions.

Hence, in one dimension, the $Neel$ state is always destabilised by quantum
corrections. This is just a manifestation of the familiar result  that 
there  is no long range order in one dimension
(Mermin-Wagner theorem) or equivalently, that there is no spontaneous symmetry 
breakdown in 1+1 dimensions 
(Coleman's theorem). Both these theorems are a consequence of the infra-red
divergences in the theory.

Other methods used in higher dimensions are fermionic and bosonic mean field
theories. By substituting ${\bf S}_i = \psi_i^\dagger {\vec \sigma}
\psi$\cite{fermimf}
or  ${\bf S}_i =  a_i^\dagger a_i$\cite{sakurai} or  
${\bf S}_i^a = i\epsilon^{abc} 
a_{ib}^\dagger a_{ic}$\cite{us}
in the Hamiltonian, we get four fermion or four boson
terms which can then be treated through appropriate mean field ansatze. But in
one dimension, fluctuations beyond the mean field
theory turn out to be infra-red divergent.
Hence, specifically in one dimension, other methods are needed. We can
divide them roughly into three categories. The first one involves the exact
solution of some model Hamiltonians by some ansatz wave-functions. For example
\begin{itemize}

\item{Heisenberg AFM for $S=1/2$}

The Heisenberg AFM for $S=1/2$ in one dimension has been solved
using Bethe ansatz\cite{bethe,indrani}. 
The solution is hard to write down, but it is known that
the ground state is unique and that there is no gap. Correlation functions
fall off algebraically. 

\item{S=1 model}

The Hamiltonian is given by
\beq
H=\sum_i {\bf S}_i \cdot {\bf S}_{i+1} - \sum_i({\bf S}_i\cdot {\bf
  S}_{i+1})^2 
\eeq
For $S=1$, this has a Bethe ansatz solution, which shows that the model has a
unique ground state with no energy gap.

\item{Models with valence bond ground states}

\bi
\item{}
The Majumdar-Ghosh Hamiltonian is given by\cite{mg}
\beq  
H=J\sum_i {\bf S}_i {\bf S}_{i+1} +J/2\sum_i {\bf S}_i {\bf S}_{i+2}.
\eeq
For $S=1/2$, the ground state is given in terms of valence bonds. There are
two degenerate ground states given by

\begin{figure}[htb]
\epsfig{figure=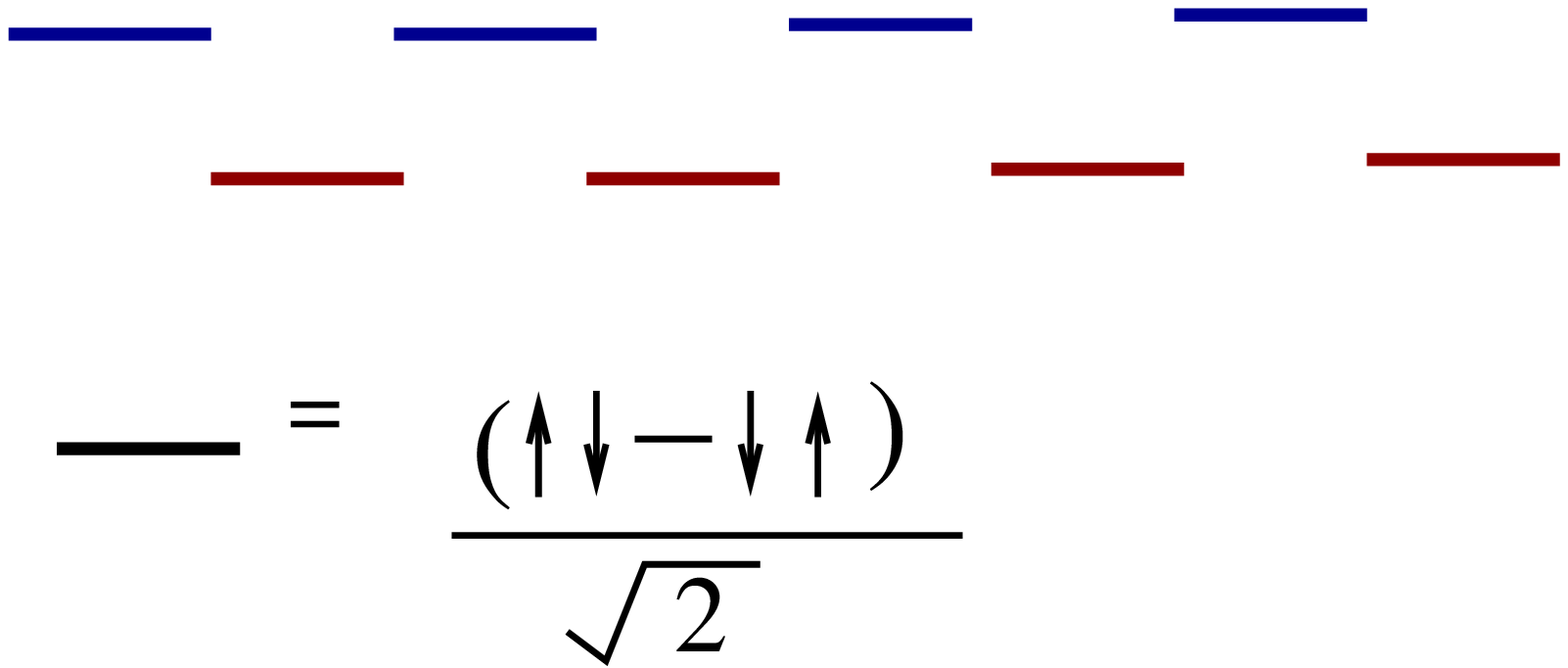,width=5.0cm}
\end{figure}

There exists a gap in the spectrum
and correlation functions have an exponential fall-off. Translational
symmetry is broken. 

\item{} The Hamiltonian for one of the  Affleck-Kennedy-Lieb-Tasaki ($AKLT$)
 models\cite{aklt}
 for $S=1$ is given by
\beq
H=J\sum_i {\bf S}_i {\bf S}_{i+1} +J/3\sum_i ({\bf S}_i {\bf S}_{i+1})^2
\eeq
This has a unique valence bond ground state found by considering each $S=1$ to
be built of a symmetrised product of 2 $S=1/2$'s.

\begin{figure}[htb]
\epsfig{figure=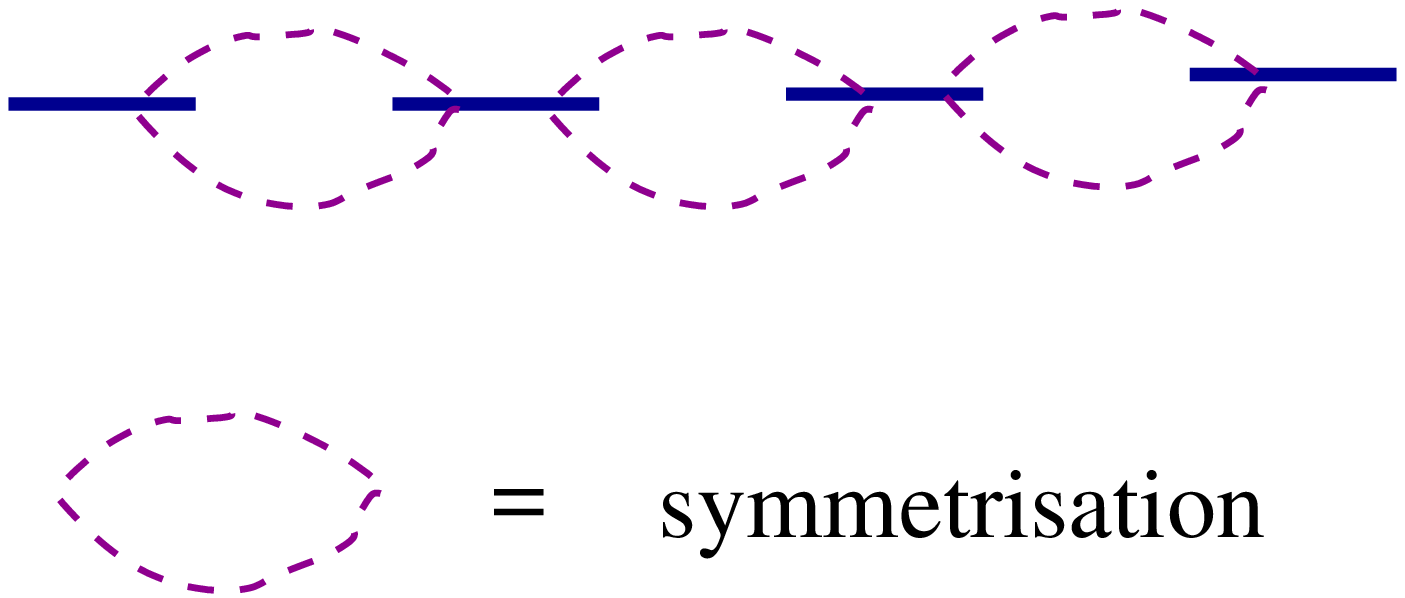,width=5.0cm}
\end{figure}
The ground state is formed by symmetrizing after forming the singlets. Here,
again, it was found that there exists a gap in the spectrum.

\ei
\end{itemize}

Besides all these explicit exact solutions of specific 
models, there is another exact
statement that can be proven in general. That is the
Lieb-Schultz-Mattis $LSM$ theorem\cite{lsm}.  
This theorem proves that the 1/2 integer spin
chain either has massless excitations or degenerate ground states
corresponding to spontaneously broken parity.

To prove this, let us start with a chain of length $L$ obeying periodic
boundary conditions. Let us call its ground state $|\psi_0>$ and assume that
this state is rotationally invariant and an even eigenstate of parity. Now
construct a new state $|\psi_1> = U|\psi_0>$ where 
\beq
U=e^{(i\pi/L) \sum_{j=-l}^l(j+l)S_j^z},
\eeq
$i.e.$, every site from $-l$ to $+l$ is rotated about the $z$ axis through
angles $i\pi/l,2i\pi/l, \dots 2i\pi$, where $l$ is some number of $O(L)$.
First, we have to show that  $|\psi_1>$ is degenerate with  $|\psi_0>$
in the $L\rightarrow \infty$ limit. To do that, we compute
\beq
<\psi_1|H-E_0|\psi_1>=  <\psi_0|U^\dagger(H-E_0)U|\psi_0> 
\eeq
where $H|\psi_0> = E_0\psi_0> $. Now using the commutation relations of the
spins, we can show that 
\beq
<\psi_1|H-E_0|\psi_1>=  {2J\pi^2\over 3 l^2} e_0(2l+2)
 \eeq
where $J$ is the coupling constant of the spins and $e_0=E_0/L$. The point to
note here is that the $R.H.S.$ is of $O(l)$ and goes to zero as
$l\rightarrow\infty$. This shows that for an infinite chain, $|\psi_0>$
and $|\psi_1>$ are degenerate. There is still a possibility that
asymptotically $|\psi_0>\rightarrow |\psi_1>$, so that we have only one state.
But to disprove that, let us look at the behaviour of $|\psi_1>$
under parity. Under parity, $S_i^z \rightarrow S_{-i}^z$ and under rotation
about the $y$-axis through $\pi$, $S_i^z \rightarrow -S_{i}^z$. Note that 
both parity
and rotation about the $y$-axis through $\pi$, are symmetries of the
Hamiltonian. Hence, under a combined action of both these symmetries,
$S_i^z \rightarrow -S_{-i}^z$. So 
\beq
U =e^{(i\pi/L) \sum_{j=-l}^l(j+l)S_j^z}\rightarrow U=e^{(-2\pi i) 
\sum_{j=-l}^lS_j^z}
\eeq
Hence, the state $|\psi_1> = U|\psi_o>$, under a combined symmetry operation
of parity and rotation, goes to $Ue^{(-2\pi i) 
\sum_{j=-l}^lS_j^z}|\psi_o> = e^{(-2\pi i) 
\sum_{j=-l}^lS_j^z}
|\psi_1>$.
But since $\sum_{j=-l}^lS_j^z=(2l+1)S$, we see that 
$e^{(i\pi/L) \sum_{j=-l}^l(j+l)S_j^z} =-1$ if the spin $S$ is odd and is
equal to +1 if
the spin $S$ is even. Hence, for 1/2 integer spins, the state $|\psi_1>$
has odd parity and is distinct from  $|\psi_0>$. In fact,
$<\psi_o|\psi_1>=0$. 
Hence, for 1/2 odd integer spins, we have proven that as $L\rightarrow
\infty$, there exists a state $|\psi_1>$ distinct from  $|\psi_0>$,
but degenerate with  $|\psi_0>$.  Hence, either there exists a massless
excitation with odd parity, or if there is a gap, then there is a degeneracy
in the spectrum. This result is  the $LSM$ theorem. The Bethe ansatz solution
for the Heisenberg AFM with massless excitations falls in the first class and
the Majumdar-Ghosh model with two degenerate ground states and massive
excitations falls in the second class. 

\subsection{2. Field theory treatment of fluctuations}

The idea here is to derive a low energy continuum limit of spin models,
keeping only the lowest derivative terms\cite{haldane,revs}. 
We shall first derive the field
theory in detail for the Heisenberg AFM, and then briefly discuss how it is
done for other general models, including the Majumdar-Ghosh model\cite{mg}.

For the Heisenberg AFM, we start by defining two fields 
\bea
{\vec \phi}_{2i+1/2} &\equiv& {\vec \phi}x_{2i+1/2} = {{\bf S}_{2i} - 
{\bf S}_{2i+1}\over 2S}, \nonumber \\
{\vec l}_{2i+1/2} &\equiv& {\vec l}x_{2i+1/2} = {{\bf S}_{2i} + 
{\bf S}_{2i+1}\over 2a}~.
\eea
Here, $a$ is the lattice spacing and the fields are defined at a point 
$x_{2i+1/2}$ between the sites $2i$ and $2i+1$ where the spins are defined.
So the pair of spin variables are now replaced by the pair of fields 
${\vec \phi}$ and ${\vec l}$. The commutation relations for the spins imply
that ${\vec \phi}(x)$ and ${\vec l}(x)$ behave like a scalar field and angular
momentum field respectively. We can also check that ${\vec\phi}^2 =
1+1/S-a^2l^2/S^2 \simeq 1$ in the large S limit. Hence, ${\vec\phi}$ is a
constrained field. 

To derive an effective field theory, we write the Hamiltonian as 
\beq
H=J\sum_{2i} [\sum_{2i} {\bf S}_{2i}\cdot  {\bf S}_{2i+1} +
{\bf S}_{2i-1}\cdot  {\bf S}_{2i} ],
\eeq
then write the spins in terms of the fields and then Taylor expand the
fields. After doing a lot of algebra, we find that
\beq
H= 2Ja \int dx [({\vec l} + \frac{S}{2}\phi')^2
+\frac{S^2\phi'^2}{4} ]
\eeq
where $\phi'= \frac{\partial{\vec\phi}}{dx}$ and $\sum_{2i}(2a) =\int dx$.
We now introduce the spin-wave volocity $v_s=2JaS$ and also the coupling
constants $g^2 =2/S$ and $\theta=2\pi S$. This allows is to rewrite the
Hamiltonian density as 
\beq
H = \frac{v}{2}[g^2(l+\frac{\theta}{4\pi}\phi')^2 +
\frac{\phi'^2}{g^2}],
\eeq which, with some more algebra can be shown to be derived from the
Lagrangian density given by
\beq
L=\frac{1}{2g^2} \partial_\mu {\vec\phi} \partial^\mu{\vec\phi}
+\frac{\theta}{8\pi}\epsilon^{\mu\nu} {\vec\phi}\cdot
\partial_\mu{\vec\phi} \times \partial_\nu{\vec\phi}
\label{o3nlsm}
\eeq
with ${\vec\phi}^2=1$. Note that we have already taken the large $S$
limit. This is necessary not only to have ${\vec\phi}^2=1$, but also to
justify the Taylor expansion. By keeping terms only upto second order in
derivatives, we are assuming that the deviations from the equilibrium
positions of the spins are small, which is justified only in the large $S$
limit. With these assumptions, we find that the spin-wave modes or
fluctuations in the $HAFM$ are described by an $O(3)$ 
non-linear sigma model ($NLSM$) with a Hopf term (the term proportional to 
$\theta$). 

The Hopf term is a total derivative, but its integral is an integer. Hence,
the action 
\beq
S = \int dt dx  L = {1\over 2g^2} \int d^2 x \partial_\mu {\vec\phi} 
\partial^\mu{\vec\phi} + i\theta Q 
\eeq
where 
\beq
Q= \frac{1}{8\pi} \int d^x \epsilon^{\mu\nu}{\vec\phi}\cdot
\partial_\mu {\vec\phi} \times  
\partial_\mu{\vec\phi}
\eeq
is an integer ( in Euclidean space). Hence, in the partition function, 
$Z=\int {\cal D}{\vec\phi}e^{-S} $, 
$e^{i\theta Q} = e^{2\pi i SQ}$ is periodic in
$S$. $S=0$ is equivalent to all $S=$ integers and $S=1/2$ is equivalent to all
$S=$ 1/2 integers.  Also, we note that for integer spins, the topological 
term can be dropped
because $^{2\pi iSQ}$ =1 for all configurations, but for half integer
spins, it is either +1 or $-1$ depending on value of $Q$. Thus the Hopf term
plays an important role for half-integer spins. This was what in fact, led to
the famous Haldane conjecture that the $HAFM$ for integer spins has a gapped
spectrum and is massless for half-integer spins. 

From the field theory mapping, in fact,it is easy to see that integer spins
models  have  a gap, but it is more non-trivial to show 
that half-integer spin models are  gapless. Let us start with a semi-classical
analysis of the integer spin models. 
Semiclassically, we assume that the $SO(3)$ symmetry of the
Lagrangian is spontaneously broken to $U(1) \simeq SO(2)$ by the Neel state or
vacuum state given by ${\vec\phi}= (0,0,1)$. Fluctuations about this state are
described by 
\beq
({\tilde \phi}_1,{\tilde\phi}_2,(1-{\tilde \phi}_1^2-{\tilde \phi}_2^2)^1/2)
\simeq ({\tilde \phi}_1,{\tilde\phi}_2,1)
\eeq
to linear order in fluctuations. Hence, the Lagrangian
\beq
L =\frac{1}{2g^2} \partial_\mu {\vec\phi} \partial^\mu{\vec\phi}
 = \frac{1}{2g^2} \partial_\mu {\vec{\tilde\phi}}_1 \partial^\mu
{\vec{\tilde\phi}}_1 + \frac{1}{2g^2} \partial_\mu {\vec{\tilde\phi}}_2 
\partial^\mu{\vec{\tilde\phi}}_2
\eeq
is just the Lagrangian of two free bosons. This is the same as the
result that was obtained using spin-wave theory.

But using the field theory, we can do a lot better. Firstly, we can use
renormalisation group (RG) to go beyond naive perturbation theory, 
$i.e.$, we can
compute the $\beta$-function. Since the manifold here ( of values taken by
the fields $(\phi_1,\phi_2,\phi_3)$) is a sphere, we can use geometric methods
to compute the RG equation and we find that
\bea
\beta (g^2) &=& \frac {dg^2}{d {\rm ln} L/a} = \frac{g^2}{2\pi}  \nonumber \\
\Longrightarrow 
 g^2_{\rm eff} (L) &=& \frac{g_0^2}{(1-({g^2_0 {\rm ln} L/a})/{2\pi})}
\eea
where $g^2$ is  the microscopic coupling that was derived at length scale $L=a$
to be $2/S$. From this, it is clear that the coupling constant blows up when 
 $({g^2_0 {\rm ln} L/a})/{2\pi}=1$ which implies $L/a = e^{2\pi/g^2} = 
e^{\pi S}$. Thus, as a function of $g^2$, we expect  a  phase transition to
the strong coupling regime, where the earlier perturbative  result of two
massless bosons is no longer valid. 
Since the  length scale  is of $O(e^{\pi S})$, masses of
order $O(e^{-\pi S})$ are expected $i.e.$, one expects to flow to a strong
coupling regime, where there is a gap of $O(e^{-\pi S})$ to excitations.

One can also substantiate this by solving the field theory in the large $N$
limit, $i.e.$, by extending the $O(3)$ $NLSM$ to $O(N)$\cite{haldane,affleck}, 
with a Lagrangian 
\beq
L = \frac{N}{2g^2} \partial_\mu {\vec\phi} \partial^\mu{\vec\phi}
\eeq
with ${\vec\phi}^2 = {\phi}_1^2+ {\phi}_2^2+ {\phi}_3^2 +\dots+\phi_N^2 =1$.
In other words, instead of having just the usual spin variables with three
components, we have extended it to $N$ components. This can also be thought of
as taking the number of dimensions in which the spin moves to be $N$.
In the limit of large $N$, it is actually possible to compute the path
integral explicitly and obtain the mass generated and we find that
\beq
m=\Lambda e^{-\pi S}
\eeq
for each of the $N$ bosons, where $\Lambda$ is an ultra-violet cutoff.
As $N \rightarrow \infty$, $S\rightarrow\infty$, but $\Lambda \rightarrow
\infty$ as well, so as to keep $m$ fixed. Higher order corrections will go as
$O(1/N)$. Having obtained this result for
large $N$, we now bravely set $N=3$ ( assuming corrections will be small) and
conclude that the integer spin HAFM has an excitation spectrum consisting of a
triplet of massive bosons with masses of the order of $e^{-\pi S}$.

All of this was for integer spins. Now what about 1/2 integer spins? Here, the
field theory includes the non-trivial Hopf term and is quite difficult to
solve. However, Affleck\cite{affleck}
 has mapped the model to a $k=1$ Wess-Zumino-Witten
($WZW$) model and by studying its symmetries, he has argued that the $\theta =
\pi$ case is massless. This difference between the integer  and half-integer
spins was the big contribution of field theories in spin models.

Similar mappings have also been used to write down field theories of other
models, such as the Majumdar-Ghosh model and its generalisations\cite{srds1,
srds2} 
For instance, for the $MG$ model for arbitary spins, we can write down an
$SO(3)_L\times SO(2)_R$ field theory\cite{srds1}
by introducing an $SO(3)$ group valued
$R$ field as follows -
\beq
 R = \left( \begin{array}{ccc} 
\phi_{11} & \phi_{21} & \phi_{31} \\
\phi_{12} & \phi_{22} & \phi_{32} \\
\phi_{13} & \phi_{23} & \phi_{33} \end{array}
\right)~.
\eeq   
In terms of the $R$ field, the Lagrangian is given by
\beq
 L = \frac {1}{4cg^2} tr ({\dot R}^T {\dot R}) - \frac{c}{2g^2}
tr (R^{'T} R' I_2)
\eeq
with $g^2$ = $\sqrt{6}/S$ and $c=JSa\sqrt{27/8}$ and $I_2$ being a
diagonal $3\times 3$ matrix with diagonal entries $(1,1,0)$ and all other
entries zero. Here, ${\dot R}$ denotes the time derviative of the matrix-valed
field $R$ and $R'$, its space derivative. The fields 
${\vec\phi}_i$ are related to the spins as 
\bea
({\vec\phi}_{1})_{3i} &=& {{\bf S}_{3i-1}-{\bf S}_{3i+1}\over \sqrt{3} S}, 
\nonumber \\
({\vec\phi}_2)_{3i} &=& {{\bf S}_{3i-1}+{\bf S}_{3i+1}-2{\bf S}_{3i}
\over \sqrt{3} S},\nonumber \\
({\vec\phi}_3)_{3i} &=&({\vec\phi}_1)_{3i} \times({\vec\phi}_2)_{3i} 
\eea
Note that the field theory has no topological term. This is not unexpected,
because here the manifold of the fields is $SO(3)$ and $\Pi_2(SO(3))=0$, 
whereas for the $HAFM$, the manifold was $S^2$ and $\Pi_2(S^2)=Z$. 
So at least naively, no
difference is expected for integer and half-integer spin models.
Also, note that the global symmetry of the action is $SO(3)_L\times SO(2)_R$,
which means that the effective action at any length scale can be written as 
\bea
L = (\frac {1}{2g_1^2}-\frac {1}{4g_2^2})
 tr ({\dot R}^T {\dot R}) &+&
(\frac {1}{2g_2^2}-\frac {1}{2g_1^2})
tr ({\dot R}^T {\dot R}I_2)\nonumber \\ +
(\frac{1}{2g_3^2}-\frac{1}{4g_4^2})tr (R^{'T} R')&+&
(\frac{1}{2g_4^2}-\frac{1}{2g_3^2})tr R^{'T} R' I_2 \nonumber
\eea
with the microscopically derived Lagrangian having $g_1^2=g_2^2=g_3^3
=2g_4^2=g^2=\sqrt{6}/S$. But these values change as we go to larger length
scales in accordance with the $RG$ equations or $\beta$-functions given by 
\bea
g_1^2 &=& {g_1^4\over 2\pi}~[~{g_1^2~g_3~g_4\over g_2^2} 
{2\over (g_1g_4+g_2g_3)} +g_1~g_3({1\over g_1^2}-{1\over g_2^2})~]~
\nonumber \\
g_2^2 &=& {g_2^4\over 2\pi}~[~g_1^3~g_3~({2\over g_1^2} -
{1\over g_2^2})^2 +2g_1~g_3({1\over g_2^2}-{1\over g_1^2})~]
\nonumber \\
g_3^2 &=& {g_3^4\over 2\pi}~[~{g_3^2~g_1~g_2\over g_4^2}
{2\over (g_1g_4~+~g_2g_3)} +g_1~g_3({1\over g_3^2}-{1\over g_4^2})~]
\nonumber \\
g_4^2 &=& {g_4^4\over 2\pi}~[~g_3^3~g_1~({2\over g_3^2} -
{1\over g_4^2})^2 +2g_1~g_3~({1\over g_4^2}-{1\over g_3^2})~]. 
\label{rgmg}
\eea
We integrated these equations numerically\cite{srds1}
and found that the length scale
where strong coupling takes over is $\zeta = L/a=e^{5.76S}$, which is of
the same order as $e^{\pi S}$ that we had found for the $HAFM$. Moreover, we
found that the flow is such that $g_1/g_2$ and $g_3/g_4$ flow to unity, so
that the symmetry gets enhanced to $SO(3)_L\times SO(3)_R$, and Lorentz
invariance is restored. Thus, the Majumdar-Ghosh model for arbitrary values of
the spin flows to a disordered phase. We shall come back to this analysis in
the last section where we study a general dimerised and frustrated model.

\subsection{3. Numerical methods}

The third method that has been used to study spin chains is through numerical
computation. Here, I shall only quote various results.

\bi

\item{}
Exact diagonalisation of small systems \\
The frustrated Heisenberg antiferromagnet modeled by
\beq
H=J[\sum_i {\bf S}_i {\bf S}_{i+1} +\alpha\sum_i {\bf S}_i {\bf S}_{i+2}]
\eeq
has been studied for $S=1/2$ to upto 20 sites and it was found that the
critical value of $\alpha$ for which a gap opens up in the spectrum
is give by
$\alpha_c=.2411\pm.0001$\cite{chitra1}. 
This is the point at which the fluid-dimer 
transition takes place.

\item{}
Density-matrix renormalisation group ($DMRG$)\cite{white} \\
This is recent method which has gained ground and is remarkably accurate.
The idea is to combine exact diagonalisation methods with the idea of
renormalisation group. So a small system is first diagonalised exactly and
then the system size is increased by adding two spins at a time on either
side. This is done repeatedly using $RG$ ideas. For the same model as above,
DMRG also finds $\alpha_c =.241$. $DMRG$ was also used to study  a more
general model involving bond alternation\cite{chitra2}.

\ei

\section{III. Frustrated and dimerised $AFM$
  spin chain}

The idea is to study the $J_1$-$J_2$-$\delta$ model given by
\beq
H=J_1\sum_i [1+(-1)^i\delta] 
{\bf S}_i {\bf S}_{i+1} + J_2 \sum_i {\bf S}_i {\bf S}_{i+2}]
\eeq 
in detail\cite{srds2}
Classically, the ground state is a coplanar configuration of the spins 
with energy per spin 
\beq
E_0 = S^2[ \frac{J_1}{2}(1+\delta) \cos \theta_1 +
\frac{J_1}{2}(1-\delta) \cos \theta_2 +J_2\cos(\theta_1+\theta_2)]
\eeq
Minimising this energy with respect to $\theta_i$ gives three phases
\bi
\item{}
Neel phase \\
\beq
\uparrow\downarrow\uparrow\downarrow \uparrow\downarrow\uparrow\downarrow \dots
\eeq
This is stable for $(1-\delta^2) > 4J_2/J_1$.

\item{}
Spiral phase \\
\begin{figure}[htb]
\epsfig{figure=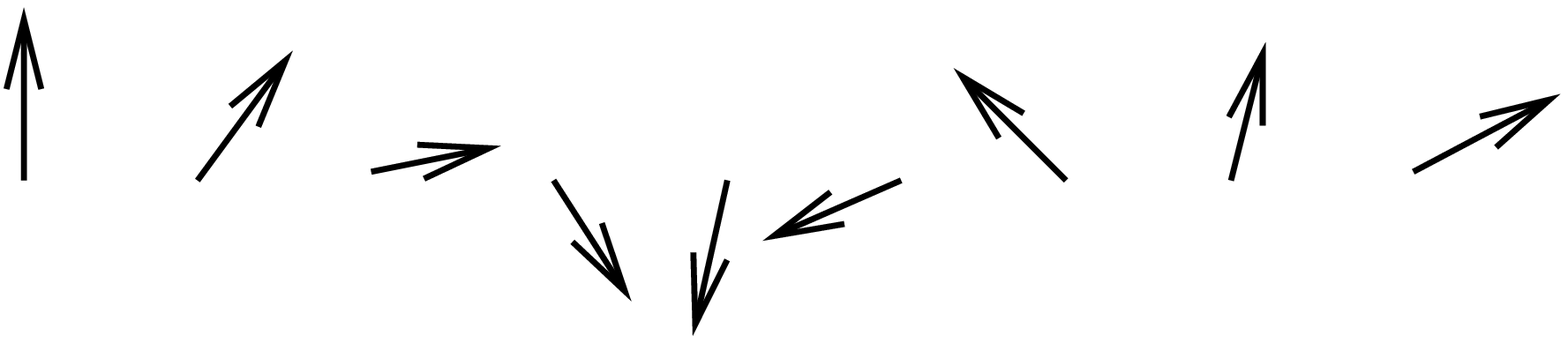,width=3.0cm}
\end{figure}
Here, the angles between neighbouring spins alternate between
$\theta_1$ and $\theta_2$ where   
\bea
\cos\theta_1 = - \frac{1}{1+\delta}[\frac{1-\delta^2}{4J_2/J_1}
+\frac{\delta}{1+\delta^2}\frac{4J_2}{J_1}] \nonumber\\
{\rm and}~~ \cos\theta_1 = - \frac{1}{1-\delta}[\frac{1-\delta^2}{4J_2/J_1}
-\frac{\delta}{1-\delta^2}\frac{4J_2}{J_1}] ~.
\eea
This phase is stable for $1-\delta^2 <4J_2/J_1 <(1-\delta^2)/\delta$.

\item{}
Collinear phase \\
This phase can be thought of as a special case of the spiral phase where
$\theta_1 = \pi$ and $\theta_2=0$. It can be denoted as
\beq
\uparrow\uparrow\downarrow\downarrow \uparrow\uparrow\downarrow\downarrow\dots
\eeq
This phase needs both frustration and dimerisation and is stable for 
$(1-\delta^2)/\delta <4J_2/J_1$.

\ei

These three phases in the classical phase diagram are depicted in Fig.(1).

\begin{figure}[htb]
\epsfig{figure=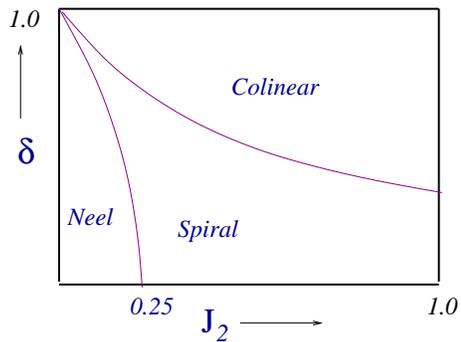,width=6cm}
\caption{Semi-classical phase diagram of the $J_1-J_2-\delta$ model}
\end{figure}

We can study fluctuations about the classical ground state as described
earlier. In the Neel phase, there are two modes with equal velocity and the
Fourier transform of the spin-spin correlation function $S(q)$ is peaked at
$q=\pi$. In the spiral phase, we have three modes, two of them with equal
velocity describe out-of-plane fluctuations and the third one with a higher
velocity describes in-plane fluctuations. $S(q)$ is peaked at
$\pi/2<q<\pi$. In the collinear phase, once again, there are two modes with
equal velocity, but here $S(q)$ is peaked at $q=\pi/2$. But as we have already
seen earlier, we do not expect spin-wave theory to be accurate in one
dimension because, there is no long-range order, no
spontaneous symmetry breakdown and no Goldstone modes in one dimension. 
   
Next, what do we know about the model exactly? For $J_2=\delta=0$, the model
is just the $HAFM$ and the solution for $S=1/2$ 
is a unique ground state with no
excitations. For $J_2=J_1/2$ and $\delta=0$, which is the $MG$ model,
the solution for $S=1/2$ is the doubly degenerate valence bond state, 
with massive
excitations. In fact, this state turns out to be the ground state even with
dimerisation along the line $2J_2+\delta=J_1$.

Now, let us study the field theory model for the fluctuations in the three
classical phases. 

\bi

\item{}
In the Neel phase, even with $J_2$ and $\delta$, the
mapping is to an O(3) $NLSM$, with the Hopf term as given in 
Eq.(\ref{o3nlsm}).
The only difference is that now $c=2J_1 aS\sqrt{1-\delta^2-4J_2/J_1}$,
$g^2=2/(S(1-\delta^2-4J_2/J_1))$
and $\theta=2\pi S(1-\delta)$. We expect the theory to have a mass gap in
general and to be massless only when $\theta=2\pi S(1-\delta) = \pi$. Note
that a topological term is present to distinguish different spins, but spin is
not really a continuous variable. So for each spin, integer or half-integer,
there are specific values of $\delta$ which can be chosen to get massless
points.   

\item{}

For the spiral phase also, the field theory still turn out to be
the $SO(3)_L\times SO(2)_R$ invariant, but with a Lagrangian given by 
\beq
 L = \frac {1}{4cg^2} tr ({\dot R}^T {\dot R}P_0) - \frac{c}{2g^2}
tr (R^{'T} R' P_1)
\eeq
where $P_0$ and $P_1$ are diagonal matrices with the diagonal elements given
by $P_0=(1/2g_2^2,1/2g_2^2,1/g_1^2-1/2g_2^2)$ and $P_1=
(1/2g_4^2,1/2g_4^2,1/g_3^2-1/2g_4^2)$ respectively.
The $RG$ equations
are the same as the ones given in Eq.(\ref{rgmg}). However, the initial
microscopic values of the coupling constants are different now and are given
by 
\bea
g_2^2&=&g_4^2~=~\frac{1}{S} \sqrt{\frac{4J_2+J_1}{4J_2-J_1}}, \nonumber \\
g_3^2 &=& 2g_2^2, \nonumber \\
{\rm and}~~g_1^2&=&g_2^2[1+(1-J_1/2J_2)^2]~.
\eea
As before, the $RG$ equations can be integrated numerically with these initial
conditions and it can be shown that the theory flows once again to an
$SO(3)_L\times SO(3)_R$ Lorentz invariant field theory.

The interesting point is that this theory turns out to be an 
exactly solved model\cite{polyakov}. 
The low energy spectrum consists of a massless spin 1/2 doublet. Hence, in the
spiral phase ( which requires sufficiently large frustration and
dimerisation), we can make the prediction that both integer and half-integer
spin models should have massive spin 1/2 excitations. The long wavelength
excitations are expected to be `two-particle' excitations, the spin triplet
and the spin singlet excitations.

Although there is no topological term in the Lagrangian, we claim that there
does exist a difference between integer and half-integer spins in this phase.
Tunneling between soliton sectors can lead to a unique ground state for
integer spins, but this is not possible for 1/2 integer spins, which have a
doubly degenerate ground state, in accordance with the $LSM$ theorem.

\item{}

Finally, we can write down the field theory for the collinear phase as
well. Here again, the field theory turns out to be an $O(3)$ $NLSM$, but
without the Hopf term. This means that the phase is always gapped both for
integer and non-integer spins. 

\ei

We generally expect these field theories to be valid in the large $S$ limit,
but for small values of $S$ such as $S=1/2$ and  $S=1$, the above analysis is
only indicative and numerical studies are needed to get the phase diagram
accurately. These have been obtained{\cite{chitra1,chitra2} and we only 
reproduce the phase diagrams here -
\begin{figure}[htb]
\epsfig{figure=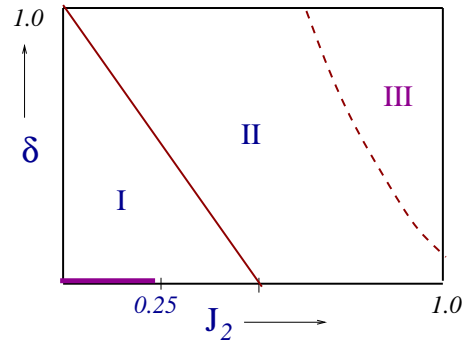,width=6cm}
\caption{$S=1/2$ phase diagram of the $J_1-J_2-\delta$ model}
\end{figure}
 \begin{figure}[htb]
\epsfig{figure=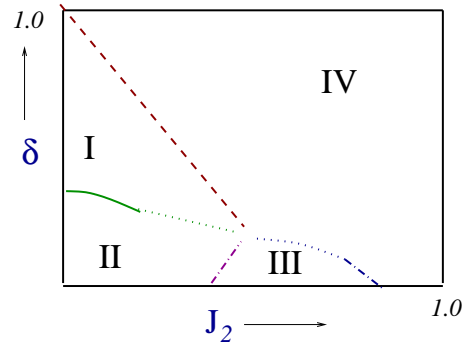,width=6cm}
\caption{$S=1$ phase diagram of the $J_1-J_2-\delta$ model}
\end{figure}

As can be seen by comparing these diagrams, with the classical phase diagram
in Fig. (1), the qualitative picture is reproduced for the spin 1/2, but for
spin $S=1$, there are many new unexpected features in the $S=1$ phase diagram
obtained numerically.

\section{IV. Conclusion}

In this paper, we have given an overview of
 the field of quantum spin chains, with
emphasis on the non-linear sigma model mapping. To recapitulate, quantum spin
 chains are spin models on a one-dimensional lattice. For parity invariant
 systems, the Lieb-Schulz-Mattis theorem says that for half-integer spin
 models, the ground state is either doubly degenerate, or the spectrum
 contains a massless mode. Using the $NLSM$ mapping, we demonstrated
that the difference between 1/2 integer spin chains and integer spin
chains was caused by the existence of a topological 
Hopf term in the Lagrangian. The presence of this term for 
1/2 integer chains led to a  gapless spectrum, whereas integer spin chains
which did not have the Hopf term were gapped.
For more general models, such as spin chains with dimerisation
and/or frustration, the $NLSM$  approach can only give a
qualitative understanding. For instance, the mapping of the Majumdar-Ghosh
model (more generally, the spiral phase of a frustrated and dimerised spin
chain to the RG fixed point Lagrangian of an $SO(3)_R\times SO(3)_L$ model 
leads to the prediction that the low energy spectrum consists of a massive
spin 1/2 doublet.    
But for low values of $S$, such as 1/2 and 1, often numerical methods
are needed to get better results, as seen in the explicit phase diagrams
 for the spin 1/2 and spin 1 frustrated and dimerised models.

One of the important issues in this field is to get a proper understanding of
the Haldane gap. Usually, a gap is formed when some symmetry is broken. So we
need a symmetry that exists for half-integer spins and is broken by all
integer spins. Since the distinction between the integer and half-integer
spins occurs because of the topological Hopf term, it is expected that the
order parameter characterising the massive and massless phases is also
topological in nature. A claim is that there exists a hidden $Z_2\times Z_2$
symmetry in the $S=1$ model, which when broken leads to the gapped Haldane
phase. But this phenomenon is not well-understood.

\vskip .5 true cm
\centerline{\bf Acknowledgments}
\vskip .5 true cm

I would like to thank Diptiman Sen for many years of useful collaboration,
including our work together in this field.

\end{document}